\documentclass[aps,prd,twocolumn,showpacs,amsmath,amssymb]{revtex4-1}
\usepackage{amsmath}
\usepackage{graphicx}
\usepackage{subfigure}
\usepackage{epstopdf}
\usepackage{color}
\usepackage{multirow}
\usepackage{setspace}
\usepackage{overpic}
\usepackage{amssymb}
\usepackage[bookmarksnumbered,pdfstartview=FitH,colorlinks,urlcolor=blue, citecolor=blue,linkcolor=blue] {hyperref}
\usepackage{lineno}
\usepackage{bm}
\usepackage{rotating}
\usepackage{verbatim}
\usepackage[utf8]{inputenc}
\usepackage{mhchem} 
\graphicspath{{Fig/}}
\usepackage{bigstrut}

\let\oldequation\equation
\let\oldendequation\endequation

\renewenvironment{equation}
  {\linenomathNonumbers\oldequation}
  {\oldendequation\endlinenomath}

\begin{document}


\title{\boldmath Search for the baryon and lepton number violating decay $J/\psi\to  pe^-$ + c.c}


\author{
M.~Ablikim$^{1}$, M.~N.~Achasov$^{4,c}$, P.~Adlarson$^{76}$, X.~C.~Ai$^{81}$, R.~Aliberti$^{35}$, A.~Amoroso$^{75A,75C}$, Q.~An$^{72,58,a}$, Y.~Bai$^{57}$, O.~Bakina$^{36}$, Y.~Ban$^{46,h}$
% ...
\\
\vspace{0.2cm}
(BESIII Collaboration)\\
\vspace{0.2cm} {\it
$^{1}$ Institute of High Energy Physics, Beijing 100049, People's Republic of China\\
% ...
$^{4}$ Budker Institute of Nuclear Physics SB RAS (BINP), Novosibirsk 630090, Russia\\
% ...
$^{35}$ Johannes Gutenberg University of Mainz, Johann-Joachim-Becher-Weg 45, D-55099 Mainz, Germany\\
$^{36}$ Joint Institute for Nuclear Research, 141980 Dubna, Moscow region, Russia\\
% ...
$^{46}$ Peking University, Beijing 100871, People's Republic of China\\
% ...
$^{57}$ Southeast University, Nanjing 211100, People's Republic of China\\
$^{58}$ State Key Laboratory of Particle Detection and Electronics, Beijing 100049, Hefei 230026, People's Republic of China\\
% ...
$^{72}$ University of Science and Technology of China, Hefei 230026, People's Republic of China\\
% ...
$^{75}$ University of Turin and INFN, (A)University of Turin, I-10125, Turin, Italy; (B)University of Eastern Piedmont, I-15121, Alessandria, Italy; (C)INFN, I-10125, Turin, Italy\\
$^{76}$ Uppsala University, Box 516, SE-75120 Uppsala, Sweden\\
% ...
$^{81}$ Zhengzhou University, Zhengzhou 450001, People's Republic of China\\
\vspace{0.2cm}
$^{a}$ Deceased\\
% ...
$^{c}$ Also at the Novosibirsk State University, Novosibirsk, 630090, Russia\\
% ...
$^{h}$ Also at State Key Laboratory of Nuclear Physics and Technology, Peking University, Beijing 100871, People's Republic of China\\
% ...
}
}

\begin{abstract}
Based on $(2712.4\pm 14.3) \times 10^{6} $ ${\psi(3686)}$ events collected by the BESIII detector operating at the BEPCII storage ring, we perform a search for the baryon- and lepton-number violating decay $J/\psi \to  pe^{-}+c.c.$ via  $\psi(3686) \to \pi^{+}\pi^{-}J/\psi$. No significant signal is found. An upper limit on the branching fraction of $\mathcal{B}(J/\psi \to p e^{-}+ c.c.) < 3.1 \times 10^{-8}$ at 90\% confidence level.
\end{abstract}

\maketitle

\section{Introduction} 
In the Standard Model, baryon number (BN) is strictly conserved. However, theoretical possibilities for baryon number violation (BNV) remain, despite the lack of experimental evidence. \cite{the8,BNV1,PDG}.
Unlike the stability of the electron, which is ensured by its status as the lightest charged particle, proton stability is tied to BN conservation, a symmetry that is not fundamental.
If BN conservation is slightly broken, as suggested by various models~\cite{the1, the2, the3, the4, the5, the6}, it could dramatically alter our understanding of the universe. BNV is crucial for explaining the current baryon-antibaryon asymmetry in the universe that began symmetrically from the Big Bang and is hence key to modern cosmology. And even a slight BNV would significantly impact the ultimate fate of the universe. 

Theoretical models suggest that BN might not be strictly conserved in nature. For instance, in Grand Unified Theory (GUT), the proton could decay into various final states via intermediate particles like leptoquarks~\cite{the8}. This could lead to BNV processes, such as $p\to e^+ \pi^0$, which would break both baryon and lepton numbers while conserving their difference, $B-L$.
Thus far, experimental constraints on proton decay have largely ruled out the simplest GUT models. Therefore, it is essential to explore other hadrons containing second-generation quarks, such as $\Lambda_{c}$ and $\Xi$, in both theory and experiment. 
Several dimension-six operators lead to BNV processes~\cite{the8, BNV1, the9}. Such operators could arise in models with R-parity-violating supersymmetric extensions of the Standard Model~\cite{R_vio} or from heavy gauge bosons like leptoquarks. 
At BESIII, several BNV processes have been investigated~\cite{BNV_total}, including $J/\psi \to \Lambda_c^+ e^-$~\cite{BES-BNV1}, $\Lambda$-$\bar\Lambda$ oscillation~\cite{Oscillation}, and various $D$ decay channels~\cite{D_meson1, D_meson2, D_meson3, BES-BNV2}. No significant signal was found, and upper limits on the branching fractions were reported.
\\

In this paper, we analyze $(2712.4\pm 14.3) \times 10^{6}$~\cite{psi_num} $\psi(3686)$ events collected by the BESIII detector~\cite{BESIII_intro} at the BEPCII storage ring~\cite{BEPCII_intro}, and search for the baryon- and lepton-number violating decay $ J/\psi \to p e^{-}$~(charge conjugation is implied throughout this paper) through the process $\psi(3686) \to \pi^{+} \pi^{-} J/\psi$.

\section{BESIII detector and Monte Carlo simulation} 
The BESIII detector is located at the south collision point of the BEPCII storage ring, which operates at center-of-mass energies between 1.85 and 4.95 ${\rm GeV}/c^2$, with a maximum luminosity $\rm 1.1 \times 10^{33} \ cm^{-2}s^{-1}$ at $\sqrt s=3.773 ~{\rm GeV}/c^2$.

The cylindrical core of the BESIII detector covers 93\% of the full solid angle and consists of a helium-based multilayer drift chamber (MDC), a plastic
scintillator time-of-flight system (TOF), and a \ce{CsI(Tl)} electromagnetic calorimeter (EMC),  all enclosed in a superconducting solenoidal magnet that provides a 1.0 T magnetic field (The magnetic field intensity in the year 2012 was 0.9 T). The solenoid is supported by an octagonal flux-return yoke, with resistive plate counter muon identifier modules interleaved with steel.

The charged-particle momentum resolution at 1 ${\rm GeV}/c$ is 0.5\%, and the specific ionization energy loss ($dE/dx$) resolution is 6\% for electrons from Bhabha scattering. The EMC measures photon energies with a resolution of 2.5\% (5\%) at 1 ${\rm GeV}$ in the barrel (end-cap) region. The time resolution in the TOF barrel region is 68 ps, while in the end-cap region, it is 110 ps. The end-cap TOF system was upgraded in 2015 using multigap resistive plate chamber technology, providing a time resolution of 60 ps~\cite{etof}. About $87\%$ of the data used in this analysis benefit from the upgrade.

Simulated samples produced with the {\sc geant4}-based~\cite{GEANT4} Monte Carlo (MC) program, which includes the geometric description~\cite{detectordescription1,detectordescription2} of the BESIII detector and the detector response, are used to determine the detection efficiency and estimate the background. The simulation incorporates the beam energy spread and initial state radiation in the $e^+e^-$ annihilations modelled with the generator {\sc kkmc}~\cite{KKMC}. The known decay modes are modelled with {\sc evtgen}~\cite{EVTGEN} using branching fractions taken from the Particle Data Group~\cite{PDG} and the remaining unknown decays from the charmonium states with {\sc lundcharm}~\cite{LUNDCHARM}. Final state radiation from charged final-state particles is incorporated with {\sc photos}~\cite{PHOTOS}. To study the detection efficiencies of the BNV decay, one million signal MC events are generated. For this analysis, 2748 million inclusive MC events including all the possible decays of the $\psi(3686)$ are used to study the background. The continuum background is estimated with the data samples at $\sqrt{s} = 3.650$ ${\rm GeV}$ and $3.682$ ${\rm GeV}$, corresponding to integrated luminosities of $410\ {\rm pb^{-1}}$ and $404\ {\rm pb^{-1}}$, respectively.

\section{DATA ANALYSIS} \label{EVTSELET}
To avoid possible bias, a blind analysis technique is employed, where the data is analyzed only after the analysis
procedure has been finalized and validated with MC simulation. 
%
%
The signal channel is $J/\psi \to p e^{-}$, via $\psi(3686) \to \pi^+\pi^-J/\psi$.  
All tracks must satisfy $|\rm{cos}\theta| < 0.93$, where $\theta$ is defined with respect to the symmetry axis of the MDC, referred to as the $z$-axis. Each track must originate from the interaction region, defined as $R_{xy} <1.0$ cm and $|R_{z}| < 10.0$ cm, where $R_{xy}$ and $R_{z}$ are the distances of the closest approach to the interaction point of the track in the $xy$-plane and $z$-direction, respectively. Events with at least four selected charged tracks and zero net charge are retained for further analysis.

%
%
To improve the detection efficiency, charged tracks with momentum less than 0.45 GeV/$c$ are assigned as pions~\cite{pipijpsi}. For particle identification (PID) of electrons and protons, the $dE/dx$, TOF and EMC information are combined to calculate the confidence levels (CL) for electron, pion, kaon, and proton hypotheses: ${\rm CL}_e$, ${\rm CL}_\pi$, ${\rm CL}_K$, and ${\rm CL}_p$. 
The electron candidates are required to satisfy ${\rm CL}_e>0.001$, ${\rm CL}_e/({\rm CL}_e+{\rm CL}_K+{\rm CL}_{\pi})>0.8$ and $0.8< E/p <1.2$, where $E$ is the energy deposited by the electron in the EMC, and $p$ is the momentum in the MDC. The proton candidates are required to satisfy ${\rm CL}_p>0.001$, ${\rm CL}_p>{\rm CL}_K$, ${\rm CL}_p>{\rm CL}_{\pi}$ and ${\rm CL}_p>{\rm CL}_e$.

%
%
In order to improve the mass resolution, a kinematic fit~\cite{fit} enforcing four-momentum conservation (4C) for four charged tracks is performed on the selected $\pi^+\pi^- pe^-$ combination. If there are multiple combinations, only the one with minimum $\chi^2_{\pi^+\pi^-pe^-,4\rm C}$ is retained. To suppress potential backgrounds, events with $\chi^2_{\pi^+\pi^-pe^-,4\rm C}<10$ are kept for further analysis. MC studies show that this requirement can veto 94\% of the background while preserving 50\% of the signal events. This requirement has been optimized using the Punzi significance method~\cite{Punzi} with the formula $\varepsilon/(1.5+\sqrt{b})$, where $\varepsilon$ is the detection efficiency from the signal MC simulation and $b$ is the number of background events from the inclusive MC sample. 

\begin{figure}[ht]
\begin{center}
\includegraphics[width=8cm]{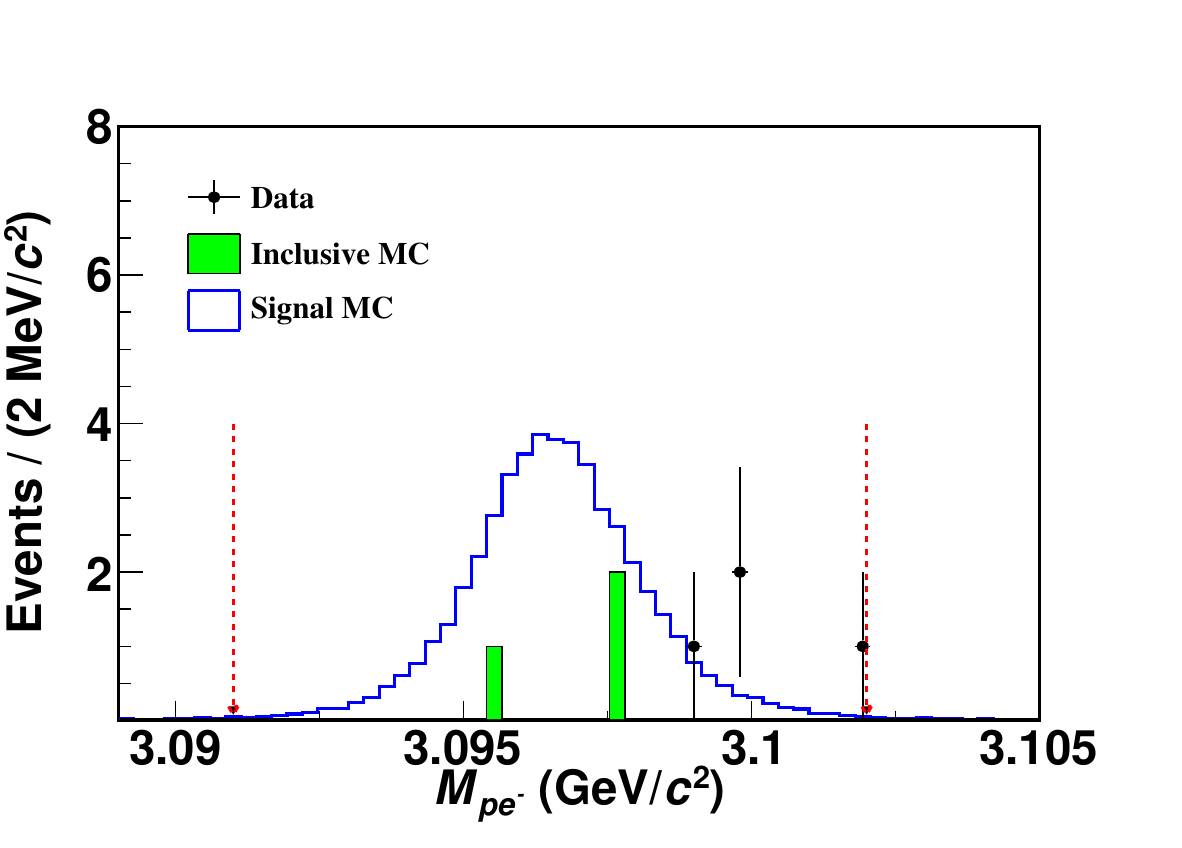}
    \caption{The $M_{pe^{-}}$ distribution shows data points with error bars, the background as a green histogram, the signal MC as a blue histogram, and the signal region is marked by red dashed arrows.} 
    \label{fig:data}
\end{center}
\end{figure}

To suppress residual background from processes with four tracks due to particle misidentification, such as $e^{+}e^{-} \to$ $\pi^+\pi^-e^{+}e^{-}$, $\pi^+\pi^-\mu^+\mu^-$, $\pi^+\pi^-\pi^+\pi^-$, $\pi^+\pi^-K^+K^-$ or $\pi^+\pi^-p\bar p$ processes, we reassign the masses of the third and fourth tracks in the $\pi^+\pi^- pe$ candidate according to the above hypotheses and perform the 4C kinematic fit again. To suppress these backgrounds, the $\chi^2_{\rm 4C}$ of each hypothesis is required to be larger than $\chi^2_{\pi^+\pi^-pe^{-},4\rm C}$.
%
%
After the above selection, the number of estimated background events based on the generic $\psi(3686)$ inclusive MC sample is found to be $N^{inc}=3$, as shown in Fig.~\ref{fig:data}. These remaining events consist of two
 $\psi(3686) \to  \pi^{+} \pi^{-} J/\psi, J/\psi \to e^{+} e^{-} \gamma^{f} \gamma^{f}$ processes and one $\psi(3686) \to  \pi^{+} \pi^{-} J/\psi, J/\psi \to e^{+} e^{-} \gamma^{f}$ process, where an $f$ superscript denotes final-state radiation.
Additionally, the continuum background, $N^{\rm cont}_{\rm bkg}$, is estimated by analyzing data samples collected at $\sqrt{s}= 3.650$ and $3.682$ GeV~\cite{N3650}. Using the same analytical procedure, no events were found within the $J/\psi$ signal window after normalization, accounting for differences in integrated luminosities and the energy-dependent cross sections as follows~\cite{ref::scale},
{\small
\begin{equation}
    N^{\rm cont}_{\rm bkg}=N_{\rm cont} \, \frac{\mathcal{L}_{3.686}}{\mathcal{L}_{\rm cont}} \, \left(\frac{E_{\rm cont}}{3.686}\right)^2,
\end{equation}
}
where $N_{\rm cont}$ represents the number of background events from the continuum process, $\mathcal{L}_{\rm cont}$ and $\mathcal{L}_{3.686}$ are the integrated luminosities for continuum and $\psi(3686)$ data, respectively, and $E_{\rm cont}$ is the center-of-mass energy for the continuum. The number of continuum background events is estimated to be $N^{\rm cont}_{\rm bkg}=0$ as $N_{\rm cont}=0$. Combining contributions,  the total background yield is estimated to be $N^{\rm bkg} = N^{\rm inc}+N^{\rm cont}_{\rm bkg} = 3$. 
%
%
Based on the fit to the distribution of the $p e^{-}$ invariant mass, $M_{pe^{-}}$, from the signal MC simulation, using a double Gaussian function and a first-order polynomial to model the signal and background shapes, respectively, the $J/\psi$ signal window  in $M_{pe^{-}}$ is set as $[3.091,~3.102]$ GeV/$c^2$, corresponding to three standard deviations.
The detection efficiency is determined to be $(19.94\pm 0.07)\%$ based on simulated $\psi(3686)\to\pi^+\pi^-J/\psi,J/\psi \to p e^{-}$ events, where $\psi(3686)\to\pi^+\pi^-J/\psi$ is generated according to the results in Ref.~\cite{ref::jpipi}, which takes into account the small dipion D-wave contribution in addition to the dominant S-wave component.

\section{Systematic Uncertainty}
Systematic uncertainties in the measurement of the branching fraction can be classified into additive and multiplicative categories. The uncertainty due to the choice of signal window is considered additive, while the sources of multiplicative uncertainties include the total number of $\psi(3686)$ events, the MDC tracking and PID efficiencies, the 4C kinematic fit, MC signal modeling, and the branching fraction for $\psi(3686) \to \pi^+\pi^- J/\psi$.
The uncertainty in the total number of $\psi(3686)$ events, determined from inclusive hadronic final states, is $0.5$\%~\cite{psi_num}.
The uncertainty in tracking charged particles is assigned as the tracking efficiency difference between data and MC simulation, evaluated using control samples of $J/\psi\to\pi^+\pi^-\pi^0$, $J/\psi\to e^+e^-(\gamma^f)$, and $J/\psi\to\pi^+\pi^-p\bar{p}$. The uncertainties are assigned as 1.0\% per pion, 1.0\% per electron, and 1.0\% per proton~\cite{BES-BNV2,MDCPID_pion}. Consequently, the total systematic uncertainty in MDC tracking is 4.0\%.
The uncertainties in PID efficiencies are estimated using high-purity control samples of electrons and protons. The resulting uncertainties are 1.1\% per electron and 2.8\% per proton~\cite{BES-BNV2,MDCPID_pion}. No PID is required for the pions, so the total systematic uncertainty in PID efficiency, conservatively allowing for full correlation, is 3.9\%, 
. 
The systematic uncertainty in 4C kinematic fit is assessed by comparing the efficiency differences before (nominal analysis) and after a tracking helix correction~\cite{Helix_method}. The relative difference in efficiencies, measured as 9.0\%, is taken as the uncertainty. 
The uncertainty in the MC modeling arises from variations in the parameters  of angular distribution, described by the formula $\frac{dN}{d\cos\theta} \propto 1+\alpha \cos^{2}\theta$, where $\theta$ is the polar angle between the beam direction and the outgoing baryon. In this work, the $J/\psi \to pe^{-}$ process is generated with $\alpha=0$, resulting in a flat angular distribution. To account for the systematic uncertainty due to MC modeling, alternative signal MC events are generated with $\alpha$ values set to be 0.68 and 1.0, respectively. The resulting standard deviation of efficiency, 0.8\%, is taken as the uncertainty.

The uncertainty in the quoted branching fraction $\mathcal B(\psi(3686)\to \pi^+\pi^-J/\psi)=(34.69\pm0.34)\%$ is 1.0\%~\cite{PDG}.
All multiplicative systematic uncertainties discussed above are summarized in Table~\ref{tab_sys}. By combining all sources in quadrature, the total systematic uncertainty is determined to be 10.7\%. 

\begin{table}[!ht]
\centering
\caption{Multiplicative systematic uncertainties in the branching fraction measurements.}
\begin{tabular}{lc}
\hline
    Source                     & Uncertainty~(\%)  \\
    \hline
    $N_{\psi(3686)}^{\rm tot}$ & 0.5 \\
    Tracking 	               & 4.0 \\
    PID 		       & 3.9 \\
    4C kinematic fit           & 9.0 \\
    MC modeling                & 0.8 \\
    ${\mathcal B}$($\psi(3686) \to \pi^+\pi ^- J/\psi$) & 1.0 \\
    \hline
    Total		       & 10.7 \\
\hline
\end{tabular} \label{tab_sys}%
\end{table}%

The uncertainty in the signal window is estimated using various ranges, from $\mu \pm 1\sigma$ to $\mu \pm 5\sigma$. Among these variations, the largest upper limit on the branching fraction of $J/\psi \to pe^-$ is chosen as our final result.
The resulting efficiency-corrected upper limit on the yield is $N^{\rm up}=29.7$, obtained using a frequentist method~\cite{TROLKE}. This estimation employs an unbounded profile likelihood approach to account for systematic uncertainties, with both signal and background events modeled by a Poisson distribution. The detection efficiency is assumed to follow a Gaussian distribution, with the systematic uncertainty is treated as the standard deviation of the efficiency. Consequently, the upper limit of $\mathcal{B}(J/\psi \to p e^-)$ at 90$\%$ CL is determined by
\begin{equation}
\mathcal{B}(J/\psi \to p e^{-})
 <\frac{N^{\rm up}}{\mathcal{B}_\psi \, {N^{\rm tot}_{\psi(3686)}}} 
 =3.1\times10^{-8},
\label{eq:BF}
\end{equation}
where $N_{\psi(3686)}^{\rm tot} = (2712.4\pm 14.3) \times 10^{6} $ is the total number of $\psi(3686)$ events, and $\mathcal{B}_\psi = \mathcal{B}(\psi(3686) \to \pi^+\pi^-J/\psi)=(34.69\pm 0.34)\%$ is the branching fraction for the decay $\psi(3686) \to \pi^+ \pi^- J/\psi$.

\section{Result}

As shown in Fig.~\ref{fig:data}, the number of observed events in the data is $N_{\rm obs}=4$ which is consistent with the estimated background yield of $N_{\rm bkg}=3$. Therefore, the upper limit on the number of signal events for $J/\psi \to p e^-$ at the 90\% CL, $N^{\rm up}$ is estimated to be 29.7. Based on equation (\ref{eq:BF}), the upper limit on the branching fraction of $J/\psi \to p e^-$ is calculated to be $3.1\times 10^{-8}$ at 90\% CL.

\section{Summary}
Using $(2712.4\pm14.3)\times10^6$ $\psi(3686)$ events collected by the BESIII detector at the BEPCII collider, we search for the baryon- and lepton-number violating decay $J/\psi \to pe^{-}+ c.c.$ for the first time. The number of observed events is consistent with the estimated background level. No obvious signals have been observed. The upper limit on the branching fraction of $J/\psi \to p e^{-}+ c.c.$ is set to be $3.1\times 10^{-8}$ at 90\% CL, which provides stronger experimental constraints compared to similar scenarios, such as $D^0\to \bar{p}e^{+}+ c.c.$~\cite{BES-BNV2} and $J/\psi\to \Lambda_c^{+}e^{-}++ c.c.$~\cite{BES-BNV1}. Our result would stimulate possible new BNV models concerning second generation quarks. This results is expected to be improved by a factor of a thousand at the next-generation super $\tau$-charm factory~\cite{ref::stcf}.

\section{Acknowledgements}
The BESIII Collaboration thanks the staff of BEPCII (https://cstr.cn/31109.02.BEPC) and the IHEP computing center for their strong support. This work is supported in part by National Key R\&D Program of China under Contracts Nos. 2023YFA1606000, 2023YFA1606704; National Natural Science Foundation of China (NSFC) under Contracts Nos. 12035009, 11635010, 11935015, 11935016, 11935018, 12025502, 12035013, 12061131003, 12192260, 12192261, 12192262, 12192263, 12192264, 12192265, 12221005, 12225509, 12235017, 12361141819; the Chinese Academy of Sciences (CAS) Large-Scale Scientific Facility Program; CAS under Contract No. YSBR-101; 100 Talents Program of CAS; The Institute of Nuclear and Particle Physics (INPAC) and Shanghai Key Laboratory for Particle Physics and Cosmology; Agencia Nacional de Investigación y Desarrollo de Chile (ANID), Chile under Contract No. ANID PIA/APOYO AFB230003; German Research Foundation DFG under Contract No. FOR5327; Istituto Nazionale di Fisica Nucleare, Italy; Knut and Alice Wallenberg Foundation under Contracts Nos. 2021.0174, 2021.0299; Ministry of Development of Turkey under Contract No. DPT2006K-120470; National Research Foundation of Korea under Contract No. NRF-2022R1A2C1092335; National Science and Technology fund of Mongolia; National Science Research and Innovation Fund (NSRF) via the Program Management Unit for Human Resources \& Institutional Development, Research and Innovation of Thailand under Contract No. B50G670107; Polish National Science Centre under Contract No. 2024/53/B/ST2/00975; Swedish Research Council under Contract No. 2019.04595; U. S. Department of Energy under Contract No. DE-FG02-05ER41374.

\end{document}